\documentclass[10pt, a4paper, twocolumn, showkeys]{revtex4}
\usepackage[utf8]{inputenc}
\usepackage[brazil]{babel}
\usepackage{epstopdf}
\usepackage{indentfirst}
\usepackage{graphicx}
\usepackage{amsmath,amssymb,amsfonts}

\makeatletter
\@ifundefined{textcolor}{}
{%
 \definecolor{BLACK}{gray}{0}
 \definecolor{WHITE}{gray}{1}
 \definecolor{RED}{rgb}{1,0,0}
 \definecolor{GREEN}{rgb}{0,1,0}
 \definecolor{BLUE}{rgb}{0,0,1}
 \definecolor{CYAN}{cmyk}{1,0,0,0}
 \definecolor{MAGENTA}{cmyk}{0,1,0,0}
 \definecolor{YELLOW}{cmyk}{0,0,1,0}
 }



\usepackage{babel}\usepackage{hyperref}

\makeatother

\usepackage{babel}

\makeatother

\usepackage{babel}

\begin{document}

\title{Magnetosphere-Ionosphere Coupling and Field-Aligned Currents

\vspace{0.7cm}

\footnotesize (Acoplamento ionosfera-magnetosfera e correntes alinhadas aos campos)}

\author{Denny M. Oliveira}

\email{dennymauricio@gmail.com}

\affiliation {EOS Space Science Center, University of New Hampshire, Durham, NH USA}

\begin{abstract}
It is presented in this paper a review of one of several interactions between the magnetosphere and the ionosphere through the field-aligned currents (FACs). Some characteristics and physical implications of the currents flowing in a plane perpendicular to the magnetic field at high latitudes are discussed. The behavior of this system as an electric circuit is explained, where momentum and energy are transferred via Poynting flux from the magnetosphere into the ionosphere.\\

\'E apresentada nesse artigo uma revis\~ao de uma entre diversas intera\c{c}\~oes entre a magnetosfera e a ionosfera atrav\'es de correntes alinhadas aos campos. Algumas caracter\'isticas das correntes fluindo em um plano perpendicular ao campo magn\'etico em altas latitudes s\~ao discutidas. O comportamento desse sistema como um circuito el\'etrico \'e explicado, onde momento e energia s\~ao transferidos via fluxo de Poynting direcionado da magnetosfera \`a ionosfera.


\end{abstract}

\maketitle

\section{Introduction}
The existence of the Earth's magnetic field was well-established long ago by the alignment of a compass needle in a particular direction. Near the Earth's surface this field behaves like a dipole field, and its existence is believed to be due to motion of electrically charged material inside the core \cite{Kivelson1996}. In the regions further away from the surface of the Earth, this field interacts with charged particles that originate on the sun, which together are known as plasma or {\it solar wind} \cite{CostaJr2011b}, \cite{Nelson2013}. The solar wind has its own magnetic field attached to it, called interplanetary magnetic field (IMF), which interacts with the Earth's dipole field. The presence of this dipole field immersed in the solar particle flow gives rise to a cavity called the magnetosphere, which is shown with some of its current systems in Fig. 1. The existence of the magnetosphere in the space environment close to the Earth is important because it shields the Earth from particles coming from different regions of space, mainly those coming from the sun. As one of the first attempts to explain such environment, in two sequential papers from 1931 about geomagnetic storms, Chapman and Ferraro \cite{Chapman1931a} came up with a theory of currents and compression of the magnetosphere, which was the first idea of a magnetopause-boundary between the magnetosphere and the ambient medium. The bow shock is formed due to the fact that the solar wind \cite{CostaJr2011a} is super-alfv\'enic\footnote{The solar wind has origin on the sun in a region called exosphere. This region is basically made up of protons (~95\%) and alpha particles (~5\%). Since the plasma there behaves like a fluid, a theory called magnetohydrodynamics (MHD) \cite{Priest} is used to study this plasma. MHD waves are generated due to oscillations of the magnetic field and ions within that region. This theory also provides a quantity to measure the propagation velocity, known as {\it Alfv\'en speed}. This plasma is then called {\it super-alfv\'enic} because its speed is greater than the Alfv\'en speed.}, and the region between the bow shock and the magnetopause is called magnetosheath. Plasma from this region has access to the ionosphere via the dayside cusps and the magnetospheric boundary layer. Descriptions of other magnetospheres, such of planets, their satellites, and even comets, can be found in \cite{Russell2001a} and \cite{Echer2010}. Therefore, the understanding of how these regions couple to each other plays an important role in studying space weather-related phenomena, such as interplanetary shocks \citep{Oliveira2013,Oliveira2014b,Oliveira2015a,Oliveira2015b,Oliveira2015c}.
\smallskip

The ``Chapman-Ferraro cavity", as the magnetosphere was referred to initially, generated interest by some researchers in the middle of the twentieth century. It was Dungey in 1961 \cite{Dungey1961}  who first proposed a model of open magnetic field lines for the magnetosphere. In this model, IMF lines reconnect with Earth's magnetic field lines, and this coupling can generate several implications in the ionosphere, such as transfer of momentum and energy from the magnetosphere into the ionosphere. In this work we review some of these implications, starting off in the ionosphere considering a partially ionized plasma. In all discussions and arguments that follow, only the open field line model of the magnetosphere will be considered.


%
%
%
%
%
%
%

\section{Current flow and conductivity in the ionosphere}

Plasmas are characterized by two classes: collisional and collisionless plasmas. In the latter, collisions are so unlikely to happen that they can be neglected in any analysis of the plasma dynamics. Collisional plasmas are generally divided in fully ionized plasma, where it has only electrons and ions, and partially ionized plasmas, which means that some neutral atoms or molecules can be found in the plasma. In this case direct collisions between charged and neutral particles dominate, while in the fully ionized case Coulomb collisions dominate \cite{Baumjohann2009}. The plasma flowing in the ionosphere is a collisional, partially ionized plasma, which can be seen in the Dungey cycle-convection scheme represented in Fig. 2. This flow drags the plasma and heats the neutral gas, connecting the E region\footnote{E region or E layer is one of the ionosphere layers composed of ionized gas lying between 90-150 $km$ above the surface of the Earth.} of the ionosphere with the magnetosphere. 
\smallskip

It is important to know how often a collision may take place in a partially ionized collisional gas, i.e., how frequent collisions between neutrals and charges happen. Assuming that neutrals, atoms or molecules, behave like heavy obstacles to the charged particles, frontal collisions may be considered with a molecular cross-section given by $\sigma_n=\pi r_0^2$, where $r_0$ is the radius of the target. If $\langle v\rangle$ represents the average velocity, the ion-neutral collision frequency is given by $\nu_{in} = n_n\sigma_n\langle v\rangle$, with the neutral particle density represented by $n_n$. Assuming that the gas is stationary in the Earth's frame, the Lorentz equation for ions with drift velocity ${\bf v}_i$ is written as

\begin{equation}
m_i\nu_{in} {\bf v}_i = e({\bf E} + {\bf v}_i\times{\bf B}) \,,
\end{equation}
where {{\bf E} and {\bf B} represent the electric and magnetic fields, $m_i$ is the ion mass, and $e$ the ion charge. In order to understand the dynamics of the ionosphere at this point, equation (1) must be solved. Magnetic field lines in high latitudes are almost parallel to each other, so ${\bf B} = B\,{\bf e}_z$ will be used for simplicity and the electric field will be assumed with components in the plane perpendicular to the magnetic field. Thus, using the ion gyrofrequency defined by $\Omega_i=eB/m_i$, the solutions for this equation are

\begin{center}
\begin{eqnarray}
\hspace{1cm}v_x&=&\frac{1 }{ 1 +\displaystyle{\left(\frac{\Omega_i}{\nu_{in}}\right)^2  }  }\left[\left(\frac{\Omega_i}{\nu_{in}}\right) \frac{E_x}{B}  + \left(\frac{\Omega_i}{\nu_{in}}\right)^2\frac{E_y}{B}     \right]  \nonumber\\
\hspace{1cm}v_y&=&\frac{1 }{ 1 + \displaystyle{ \left(\frac{\Omega_i}{\nu_{in}}\right)^2    }}\left[ \left(\frac{\Omega_i}{\nu_{in}}\right) \frac{E_y}{B}  - \left(\frac{\Omega_i}{\nu_{in}}\right)^2\frac{E_x}{B}     \right] \nonumber
\end{eqnarray}
\end{center}

These solutions can be written in a compact, vectorial form that represents the velocity perpendicular to the magnetic field:
\begin{equation}
{\bf v_\perp} = \frac{1 }{ 1 + \displaystyle{\left(\frac{\nu_{in}}{\Omega_i}\right)^2}}    \left[ \left(\frac{\nu_{in}}{\Omega_i}\right)\frac{{\bf E}}{B} + \frac{{\bf E}\times{\bf B}}{B^2}      \right]
\end{equation}

It is interesting to analyse this equation in terms of the ratio $\nu_{in}/\Omega_i$ because the ion motion changes with it. At typical heights in the ionosphere, 100 $km$, the gyrofrequency does not change in an appreciable way, as was already discussed the magnetic field is considered practically constant over ionospheric altitudes\footnote{The Earth's magnetic field in the radial direction has strength\\
$B(r) = B_0\left(\frac{R_E}{r}\right)^3$, where $B_0$ is the magnitude of the magnetic field on the Earth's surface, $R_E$ is the Earth radius, and $r$ the distance from the Earth's center. Considering that the Earth's radius is approximately 6400 $km$, a distance variation inside the range of 100 $km$ may be neglected. }. What changes drastically is the ion-neutral collision frequency, as a result of the decreasing ion density with increasing altitude. At approximately 125 $km$ of height above the sea level, this ratio is taken as reference with $\nu_{in}/\Omega_i =  1$. Above this height, collisions are so rare that ions move mostly under ${\bf E}\times{\bf B}$ drift. Below 125 $km$, collisions happen with larger frequency and the movement of ions through the electric field dominates. This response to the neutral density in the ionosphere suggests that there are currents that obey a certain pattern, as we shall discuss in the sequence.

\begin{center}
\begin{figure}[h]
\includegraphics[scale=0.55]{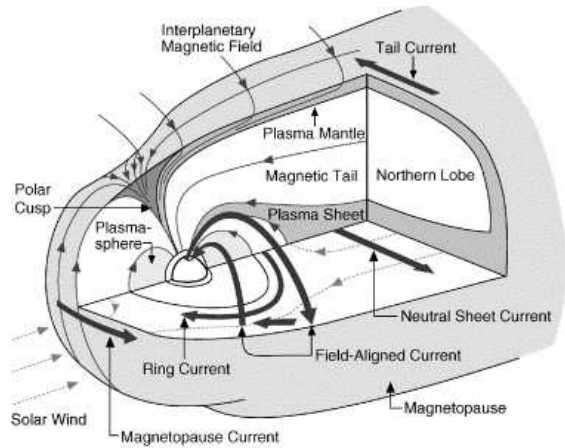}
\caption{The magnetosphere and some of its current systems, as seen in {\it Russell}, 2001. Our interest is centered in the implications generated by the field-aligned currents in the magnetosphere-ionosphere coupling.}
\label{figure}
\end{figure}
\end{center}

The movement of electrons in the ionosphere can be discussed in the same way. It is straightforward to write an expression for the current density in the ionosphere. Since the electron velocity in terms of the current density is represented by ${\bf v}_e = -1/(en_e)\,{\bf j}$, and the electron gyrofrequency has the opposite sign when compared to the ion gyrofrequency, we may define the current density perpendicular to the magnetic field solving equation (1) again:
 
 \begin{equation}
{\bf j_\perp} = \frac{en_e\displaystyle{\left(\frac{\nu_{en}}{\Omega_e}\right)} }{ 1 + \displaystyle{\left(\frac{\nu_{en}}{\Omega_e}\right)^2}}    \left[ \frac{{\bf E}}{B} -\displaystyle{\left(\frac{\nu_{en}}{\Omega_e}\right)} \frac{{\bf E}\times{\bf B}}{B^2}      \right]\,,
\end{equation}
where $n_e$ is the electron density. It is useful to redefine equation (3) in terms of the {\it Pedersen} (proportional to $\sigma_P$) and {\it Hall} (proportional to $\sigma_H$) currents:

\begin{equation}
{\bf j_\perp} = \sigma_P{\bf E} + \sigma_H\frac{{\bf E}\times{\bf B}}{B}
\end{equation}

In order to describe in which regions these currents exist, it is necessary to analyse equation (3). The Pedersen current flows parallel to the electric field at altitudes above 125 $km$. Although the ions have dominant mobility parallel to ${\bf E}$, some of them drift in the direction of ${\bf E}\times{\bf B}$. The Hall current is located at lower heights, where it flows in the direction opposite to  the electron ${\bf E}\times{\bf B}$ drift. When the ratio between the collision frequency and the electron gyrofrequency is close to unity, both currents have approximately the same magnitude. Fig. 3 shows Pedersen and Hall currents plotted in terms of $en_eE/B$ against the ratio $\Omega_e/\nu_{en}$. 
\smallskip

For future use, it is important to represent the current density in terms of the total height-integrated current perpendicular to the magnetic field:

\begin{equation}
{\bf i}_\perp = \Sigma_P{\bf E} + \Sigma_H\frac{{\bf E}\times{\bf B}}{B}\,,
\end{equation}
where $\Sigma_{P,H}=\int\sigma_{P,H}dz$ are the height-integrated Pedersen and Hall conductivity.  On the dayside of the ionosphere, these conductivities are of the order of 10 mho, which is the unit that represents the inverse of the unit of electric resistance ohm commonly used in plasma physics.

\begin{center}
\begin{figure}[h]
\includegraphics[scale=0.57]{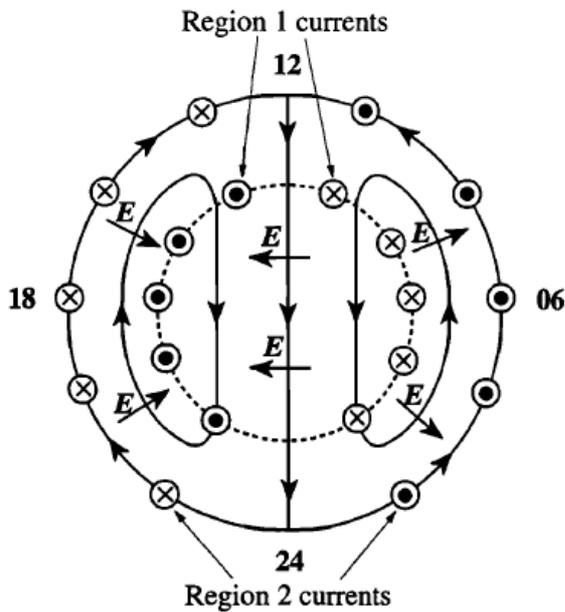}
\caption{Schematic representation of Dungey's cycle flow in the ionosphere in a noon-midnight diagram. Figure extracted from  {\it Cowley}, 2000 \cite{Cowley2000}. Dotted (away from the ionosphere) and crossed (into the ionosphere) circles represent the magnetic field lines. Electric fields are already shown in the figure.}
\label{figure}
\end{figure}
\end{center}
\vspace{-0.4cm}

The currents driven by the magnetosphere combine with the magnetic field generating the force ${\bf j}\times{\bf B}$. These electromagnetic forces compensate the drag force of the neutral particles in the ionosphere, i.e., component of ${\bf j}\times{\bf B}$ associated with the Pedersen current has direction opposite to the atmospheric drift force of the neutral particles. Similarly, the component of the same force associated with the Hall current is directed oppositely to the electric field. Therefore, each Pedersen/Hall component of ${\bf j}\times{\bf B}$ has direction pointing to the Hall/Pedersen direction, respectively. Assuming that those currents have the same order of magnitude, it may be inferred that they balance each other. Details of how the ${\bf j}\times{\bf B}$ force balances the neutral drag force and how it connects with the magnetosphere may be found in the literature \citep{Cowley2000,Kelley2009}. The height-integrated Joule heating rate per unit area is represented by ${\bf i}_\perp\cdot {\bf E}=\Sigma_PE^2$ $Wm^{-2}$, which suggests that the Pedersen current behaves like a load and the Hall current does not dissipate energy in the ionosphere.

\begin{center}
\begin{figure}[h]
\includegraphics[scale=1.02]{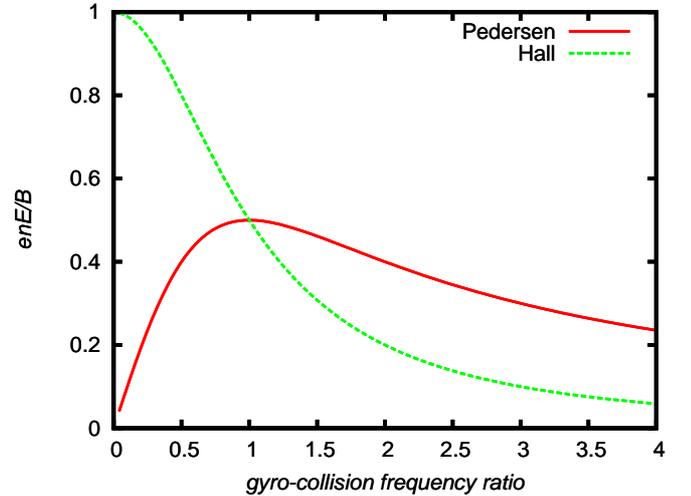}
\caption{Pedersen (red) and Hall (green) current densities plotted in terms of the inverse of the collision frequency and gyrofrequency ratio, $\Omega_e/\nu_{en}$, which increases with the height. At an altitude close to 125 $km$, the ratio approaches one and both currents have approximately  the same magnitude. Below this reference height, collisions are very likely to occur and the Hall current density dominates. Finally, above this height, collisions are so rare meaning that the ratio in Equation (3) is almost null. Therefore, in those regions, the Pedersen current density dominates.}
\label{figure}
\end{figure}
\end{center}

These currents, well known as {\it large-scale currents}, are depicted in Fig. 2 and Fig. 4. Also represented in Fig. 4 are other important currents, called field-aligned currents (FACs). Their importance and dynamic relations with both Pedersen and Hall currents will be discussed in the following section.

\section{Current circuits in the magnetosphere-ionosphere system}

The system of currents shown in Fig. 2 is based on Dungey's idea of open field lines in the magnetosphere. In that diagram the arrowed lines represent the plasma flux in the ionosphere, the dashed lines represent the boundary between open and closed field lines, and the short arrows represent the electric field. The dotted and crossed circles indicate the flow direction of the FACs, where the former indicate currents upward and the latter currents downward in relation to the ionosphere.
\smallskip

As we have derived previously, the Pedersen current always flows in the direction parallel to the electric field and, therefore, cannot close in the ionosphere. In order to satisfy the divergence of the current density, some large-scale currents must exist. These currents are called Region 1 and Region 2 currents, and are also classified as FACs. Region 1 currents are located in regions of higher latitudes, while Region 2 currents are  found in regions of lower latitudes. Region 1 currents are directed toward the ionosphere on the dawn side, and the Region 2 currents flow away from the ionosphere in the same region. On the other hand, on the dusk side the situation is the opposite: Region 1 currents flow upward and Region 2 currents downward. For a detailed discussion on how these currents close in the magnetosphere \citep{Kelley2009}. These field-aligned currents were first observed in 1966 by  Zmuda et. al. \citep{Zmuda1966}, based on a series of measurements of magnetic disturbances. Ten years later,  Iijima and Potemra \citep{Iijima1976}
and in other future works, based on a series of measurements of magnetic disturbances, found that the Regions 1 and 2 currents obey an almost permanent pattern in the ionosphere-magnetosphere system. 

\begin{center}
\begin{figure}[h]
\includegraphics[scale=0.25]{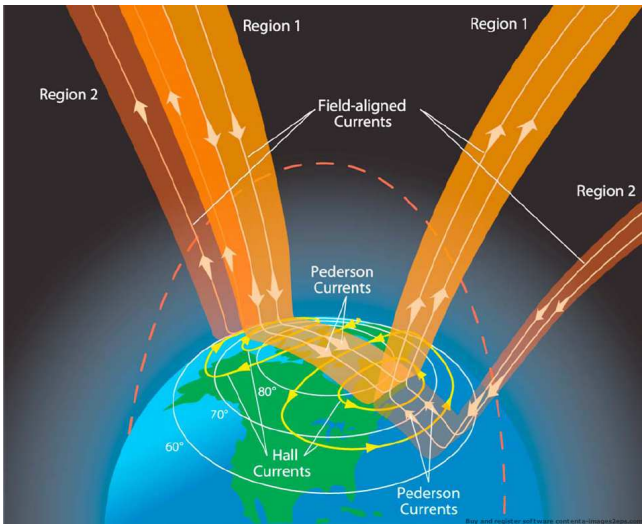}
\caption{Schematic representation of the large-scale currents in the region near the polar cap. Perdersen and Hall currents, as well as the Region 1 and Region 2 currents, are shown. Image downloaded from \url{http://commons.wikimedia.org}.}
\label{figure}
\end{figure}
\end{center}

As we have seen in the discussion of currents and conductivities in the ionosphere, the Pedersen conductivity may be thought as working like a``load" in an electric circuit. This circuit is coupled to the magnetosphere through the FACs, and the solar wind is then known as an MHD generator, which is the voltage source for such circuit. In order to understand how the magnetosphere is linked to the MHD generator, the scalar product ${\bf j}\cdot{\bf E}<0$ indicates  the source (generator), while in the low ionosphere the Pedersen currents are the load with ${\bf j}\cdot{\bf E}>0$. The flow of energy and momentum takes place due to the existence of a perturbed magnetic field which points in the direction of the sun. This perturbation is located in the direction midnight-noon in Fig. 2. As a result, a Poynting flux pointing downward arises due to the interaction of this perturbed magnetic and the electric fields, which is oriented from dawn to dusk.

\section{Conclusion}

The discussion of the coupling between the magnetosphere and the ionosphere was based on Dungey's model of open magnetic field lines. This condition then implies that there is a flux of both momentum and energy driven by the solar wind MHD generator in the magnetosphere into the ionosphere, that behaves like a load in a coupled electric circuit. This discussion was motivated by the model based on interaction among several currents located in the ionosphere in low altitude, and how they are closed with the magnetosphere through field-aligned currents (FACs). Due to the fact that electricity flows easily through the field lines in a plasma, the argument that the transfer of energy and momentum takes place because of the existence of such currents was supported. The reason for the existence of these currents is that the resistive behaviour of the ionosphere requires FACs to close the divergence of the current density. Then, if the conductivity in the ionosphere were constant, FACs would not exist and the coupling between the magnetosphere and the ionosphere would not take place.

\begin{acknowledgments}
This work was supported by grant NNX13AK31G from NASA, grant AGS-1143895 from the National
Science Foundation, and grant FA-9550-120264 from the Air Force Office of Sponsored Research.
The author is pleased to thank Dr. Charlie Farrugia for suggesting this subject and for his valuable discussions. The author is also grateful to Dr. Eberhard M\"obius and Ian Cohen for reading the manuscript and giving important suggestions.
\end{acknowledgments}


\end{document}